\input epsf
\documentstyle[preprint,aps]{revtex} 
\tighten
\begin{document}
\draft
\preprint{\vbox{Submitted to Physical Review Letters 
		          \hfill FSU-SCRI-97-080 \\
		          \null\hfill CU-NPL-1150}}
\title{Dynamic Spin Response for Heisenberg Ladders}
\author{J. Piekarewicz}
\address{Supercomputer Computations Research Institute, \\
         Florida State University, Tallahassee, FL 32306}
\author{J.R. Shepard}
\address{Department of Physics, \\
         University of Colorado, Boulder, CO 80309}
\date{\today}
\maketitle
 
\begin{abstract}{\rm }
We employ the recently proposed plaquette basis to investigate static 
and dynamic properties of isotropic 2-leg Heisenberg spin ladders. 
Simple non-interacting multi-plaquette states provide a remarkably 
accurate picture of the energy/site and dynamic spin response of 
these systems. Insights afforded by this simple picture suggest a 
very efficient truncation scheme for more precise calculations. 
When the small truncation errors are accounted for using recently 
developed Contractor Renormalization techniques, very accurate 
results requiring a small fraction of the computational effort of 
exact calculations are obtained. These methods allow us to determine 
the energy/site, gap, and spin response of $2 \times 16$ ladders. 
The former two values are in good agreement with density matrix
renormalization group results. The spin response calculations show 
that nearly all the strength is concentrated in the lowest triplet
level and that coherent many-body effects enhance the response/site by 
nearly a factor of 1.6 over that found for $2 \times 2$ systems.
\end{abstract}
\pacs{PACS numbers:~75.10.Jm,~75.40.Gb,~75.40.Mg}

\narrowtext

\vskip 0.3 true in

 Heisenberg spin ladders are deceptively simple systems consisting of 
$n_{c}$ parallel spin chains (legs) connected by pairwise interactions 
(rungs). Topologically, ladders are intermediate to chains 
($n_c\rightarrow 1$) and planes ($n_c\rightarrow \infty$) but, as is 
by now well-established, they possess physical properties which cannot 
be guessed on the basis of their behavior in the 1D and 2D limits. 
Perhaps the greatest surprise is the existence of a spin gap for 
even-leg antiferromagnetic spin-$1/2$ ladders; chains and planes are 
gapless as are odd-leg ladders~\cite{dagric96}. Apart from their 
intrinsic interest as rich mathematical systems, spin ladders are 
currently the objects of intense scrutiny because of the possibility 
that they may afford insights into the dynamics of high temperature 
superconductors~\cite{dagric96,dag92}. 
Such inquiries are enlivened by the fact that spin 
ladders appear to be realized in certain compounds, 2-leg ladders in 
the form of vanadyl pyrophosphate being perhaps the most prominent 
example~\cite{dagric96,john87}. During the past few years, consistent 
theoretical and 
experimental pictures of the static properties of spin ladders 
({\it e.g.}, the energy/site and the gap) have emerged. On the 
theoretical side, a variety of numerical methods have been employed 
including direct diagonalization, Lanczos techniques, quantum Monte 
Carlo (QMC)~\cite{barn93} and density matrix renormalization group (DMRG) 
approaches~\cite{white94}.
Progress in understanding dynamical features---including spin-triplet 
dispersion relations and spin responses~\cite{barn93}---has been somewhat 
slower, 
in part because the most powerful methods for studying static quantities, 
such as QMC and DMRG, cannot straightforwardly provide dynamical 
information. Studies of dynamical properties have typically relied on 
a combination of brute-force diagonalization (aided by Lanczos)~\cite{barn93}
and analytic (or semi-analytic) methods based on perturbation 
theory~\cite{barn93,white94,sacbha90,srz94}, the latter frequently 
being used qualitatively to illuminate the underlying physics of the 
former. In the present letter, we suggest that such studies are hindered 
by two shortcomings: 
({\it i}) that numerical and analytic treatments usually 
employ different bases and ({\it ii}) that neither basis is ideally 
suited to isotropic 2-leg spin ladders. In a recent paper~\cite{piek96} 
we proposed an alternative basis---to which we refer as the ``plaquette
basis''---which is well-suited to both numerical and analytic treatments 
and which reveals the physical origin of many dynamical features in an 
especially clear fashion. In the present paper, we exploit the plaquette 
basis to compute static and dynamic properties of 2-leg ladders with up 
to 32 sites. The results of the full brute-force diagonalizations 
(which are of course independent of the basis) can be reproduced with 
remarkable accuracy by an almost trivial model of non-interacting 
plaquettes. 

 The conventional ``$S_z$'' basis is very simple to construct and has 
been used almost exclusively in direct diagonalization calculations. 
A basis state is simply the product of spin-up or spin-down wavefunctions 
at each site. Matrix elements are simple to evaluate but, as shown in 
Ref.~\cite{piek96,dag97}, eigenvectors are very complicated. 
Analytic methods typically employ the ``rung'' or 
``singlet-triplet'' basis in which the Hamiltonian is diagonal 
in the limit in which the coupling along the rungs $(J_{\perp})$ 
is much stronger than the coupling along the chains 
$(J_{\parallel})$~\cite{barn93}. 
In this limit, the system consists of independent rungs, each of 
which can have overall spins of zero or one with eigenenergies of 
$-3/4\ J_\perp$ and $+1/4\ J_\perp$, respectively. The interaction 
proportional to $J_\parallel$ is treated as a perturbation---which 
is problematic for the physically relevant~\cite{dagric96} case of 
$J_\parallel\simeq J_\perp$. This basis has not been used in 
direct diagonalization calculations until very recently ~\cite{dag97}.

The plaquette basis~\cite{piek96} is more complicated to construct. 
The ladder is decomposed into distinct pairs of 
adjacent rungs. These $2\times2$ objects are the plaquettes. 
For the isotropic ($J_\parallel=J_\perp \equiv 1$)
two-leg ladders studied here, the Hamiltonian is 
\begin{equation}
   H=\sum_{\langle i,j \rangle}{\bf S}_{i}\cdot{\bf S}_{j} \;
\end{equation}
where the sum is over nearest-neighbor pairs.
This Hamiltonian can be written as the sum of $H_0$, which contains the 
intra-plaquette interactions, and $V$, which includes the interactions 
between plaquettes. The intra-plaquette Hamiltonian $H_0$ is expressible 
as 
\begin{equation}
  H_0=\sum_i\ h^{(i)}_0 \;;\quad 
  h^{(i)}_0 = (\vec S^{(i)}_1 + \vec S^{(i)}_4)\cdot
              (\vec S^{(i)}_2 + \vec S^{(i)}_3) \;,
  \label{hzero} 
\end{equation}
where $i$ labels a specific plaquette. 
$H_0$ is diagonal in the plaquette basis: 
\begin{equation}
  H_0|\Phi_{\rm plaq}\rangle=
  E_{\rm plaq} |\Phi_{\rm plaq}\rangle \;,
  \label{plaquettebasis} 
\end{equation}
where $|\Phi_{\rm plaq}\rangle\;(E_{\rm plaq})$ is a direct product
(sum) of single-plaquette wavefunctions (energies), each of the 
form~\cite{piek96}:
\begin{eqnarray}
  |\phi_{\alpha}\rangle &=& 
  |(s_{1}s_{4})l_{14},(s_{2}s_{3})l_{23};jm\rangle \;, \\
  \epsilon_{\alpha}&=&{1\over 2}
  \bigl[j(j+1)-\ell_{14}(\ell_{14}+1)-\ell_{23}(\ell_{23}+1)\bigr] \;.
  \label{oneplaquette} 
\end{eqnarray}
In this basis the two {\it diagonal} pairs of spins are coupled to 
well-defined total angular momentum, $\ell_{14}$ and $\ell_{23}$, 
which can equal zero or one. These two link angular momenta are in 
turn coupled to a total plaquette angular momentum $j$ with projection 
$m$. 

The single plaquette ground state has $\ell_{14}=\ell_{23}=1$ and $j=0$;
the energy of this state is $-2$. The next lowest (triply degenerate) 
state also has $\ell_{14}=\ell_{23}=1$ but $j=1$; its energy is $-1$. 
The remaining states have energies greater than or equal to zero. Thus, 
the low-energy degrees of freedom for a single plaquette always have 
diagonal or link spins coupled to triplets and total plaquette angular 
momentum 
of zero or one. In this sense, as far as low-energy configurations are 
concerned, the intra-plaquette interactions ``freeze out'' all but 
triplet-triplet modes. 
These observations suggest that the low-energy spectrum of 
multi-plaquette systems will be relatively simple to describe
in the plaquette basis where the relatively few ``frozen'' 
configurations---made up of only triplet-triplet single-plaquette 
states---should dominate. Examination of exact eigenstates determined
by direct diagonalization confirms this speculation. 
For example, the singlet ground state of the 
$2\times 8$ ladder is dominated by the one frozen configuration 
corresponding to the non-interacting ($V\rightarrow 0$) ground state 
({\it i.e.}, four frozen $j=0$ single-plaquette states). This 
particular configuration appears with an amplitude of 0.776. 
Moreover, four frozen configurations---out of a total of 1430 
states in the basis---account for 82\% of the total probability. 
We also note that simple non-interacting multi-plaquette states 
yield a surprisingly good estimate of the energy/site, namely, 
$-0.5$, which compares reasonably well with the bulk value as 
determined by DMRG~\cite{white94}, namely, $-0.57804$.
Simple second order perturbation theory~\cite{dag97} improves the 
plaquette basis value to $-0.55816$ which corresponds to a 3.4\% 
discrepancy.
The price which must be paid for using the plaquette basis is that 
matrix elements are not simple to evaluate but can be handled with 
the sophisticated methods of Racah algebra used routinely in atomic 
and nuclear physics~\cite{angmom}.

 The dominance of frozen configurations suggests that a truncation 
which retains only these states should be quite accurate. As shown 
in Ref.~\cite{piek96}, such is indeed the case. Energies for 
low-lying states of $2 \times 6$ ladders computed using this truncated 
basis (to which we refer as the ``frozen basis'' in what follows) are 
within 5 to 10\% of the exact values. As the frozen basis is much 
smaller than the full basis, its use results in a vast reduction of 
computational effort. As was also discussed in Ref.~\cite{piek96}, 
it is possible to systematically correct for these small truncation 
errors by constructing an effective low-energy Hamiltonian to be used 
with the frozen basis. We accomplish this via the recently developed 
COntractor REnormalization (CORE) method~\cite{morwei94}. Here the 
effective Hamiltonian is expressed as a cluster expansion. In the CORE 
lexicon, truncations induce new ``range-$r$'' interactions which depend 
on the quantum numbers of clusters containing $r$ adjacent blocks; in 
our present treatment, these blocks are the plaquettes. 
In Ref.~\cite{piek96} we showed that, for $2 \times 6$ ladders,  
including range-2 CORE contributions for calculations of energies of 
low-lying states in the frozen basis brought discrepancies with 
exact results down to the level of 2\% or less.

 In the present work, we examine larger systems, the largest being a 
$2 \times 16$ ladder. Moreover, we include range-3 contributions to 
the effective Hamiltonian. At this order in the CORE treatment, the 
low-energy spectrum of
the $2 \times 6$ ladder becomes exact by construction. The prediction
for the ground state of the $2 \times 8$ ladder in the frozen basis 
at range-3 differs from the exact result by less than 0.02\%; the 
discrepancy for the first excited state is even smaller. Our present 
calculations of the energy/site and the gap are summarized in 
Table ~\ref{tableone}. We compare our range-3 results with either
exact (for $2 \times 8$) or DMRG~\cite{dmrg97} (for $2 \times 16$) 
calculations. 
We also note that the CORE calculations for the $2 \times 16$ ladder 
require roughly the same computational effort as the exact calculations 
for the $2 \times 8$ system. 

 We now turn to the computation of dynamical quantities, focusing 
on the dynamical spin response,
\begin{equation}
  S(\vec q,\omega)=\sum_n\ 
  |\langle \Psi_{n}|\vec{S}(\vec q)|\Psi_{0}\rangle|^2\ \ 
  \delta (\omega - \omega_n) \;,
  \label{spinresponse} 
\end{equation}
where $|\Psi_{0}\rangle$ is the exact ground state of the system
and $|\Psi_{n}\rangle$ is an excited state with excitation energy
$\omega_n$. Since the transition operator $\vec{S}(\vec q)$ transforms
as a rank-1 tensor ({\it i.e.}, a vector) the only excited states
than can be reached from the singlet ground state have total angular
momentum of one. The spin transition operator is 
\begin{equation}
  \vec S(\vec q)=\sum_i\ \vec S_i\ e^{i\vec q\cdot\vec r_i} \;,
  \label{spinoperator} 
\end{equation}
where the sum is over all sites and $\vec q$ is the momentum transfer 
to the ladder. 
$S(\vec q,\omega)$ can 
be measured, {\it e.g.}, by inelastic neutron scattering~\cite{ecc94}. 
Of particular interest is the dynamic spin response probed at a momentum
transfer (in units of the lattice spacing) of 
$\vec q \equiv \vec q_{\pi\pi}=(\pi,\pi)$. Clearly
\begin{equation}
  \vec S(\vec q_{\pi\pi})= \sum_i\ (-)^{\langle i\rangle}\vec S_i \;,
  \label{spipip} 
\end{equation}
where $(-)^{\langle i\rangle}$ is a phase which flips in going from 
a site to any of its nearest neighbors. The essential properties of 
$S(\vec q_{\pi\pi},\omega)$ are readily apparent in the plaquette basis. 
For a single $2 \times 2$ plaquette,
\begin{equation}
  \vec S(\vec q_{\pi\pi}) \rightarrow 
        (\vec S_1 + \vec S_4) - (\vec S_2 + \vec S_3) =
         \vec L_{14} - \vec L_{23} \;.
  \label{spipipp} 
\end{equation}
We see immediately that this operator {\it cannot} break the frozen
triplets. Hence the only matrix element we need to consider connects 
the $\ell_{14}=\ell_{23}=1$, $j=0$ ground state to the 
$\ell_{14}=\ell_{23}=j=1$ first excited state. The strength/site of 
this response is 2/3. No other states are excited. Identical results 
are found, of course, for larger systems in the limit of non-interacting 
plaquettes. Exact calculations of this response for $2 \times L$ ladders 
with $L=2,4,6$ and $8$ are remarkably consistent with this simple
picture; a large fraction---typically in excess of 96\%---of the total
$S(\vec q_{\pi\pi},\omega)$ strength is concentrated in the lowest
triplet level. The response/site to this level grows slowly but steadily 
from a value of 2/3 for $L=2$ to 0.99934 for $L=8$. We again emphasize 
the role of the plaquette basis in understanding the basic features of
$S(\vec q_{\pi\pi},\omega)$.

\begin{figure}[h]
 \null
 \vskip0.8in
 \includegraphics{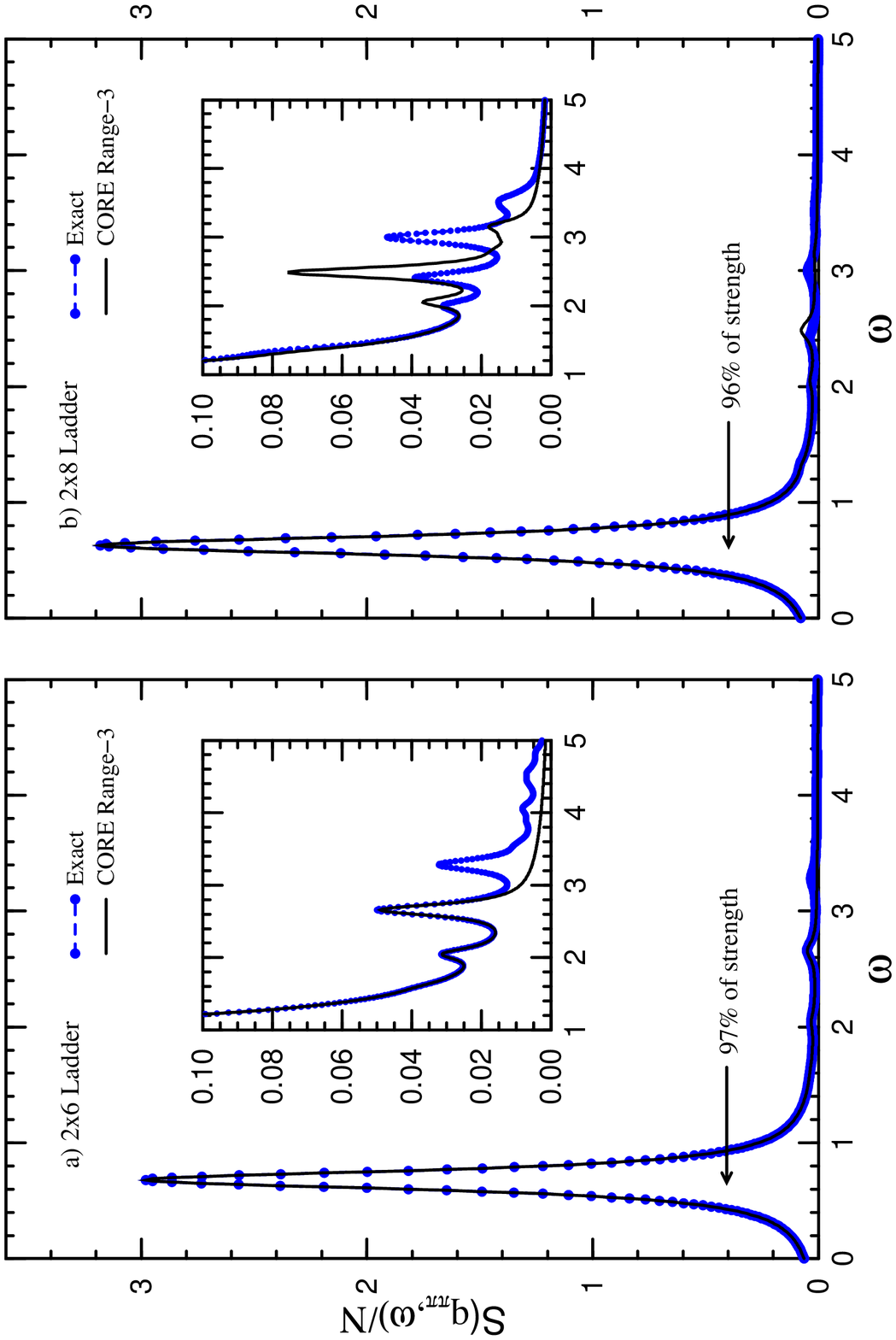}
 \vskip2.6in
\caption{Dynamic spin response for a) $2\times 6$ and 
	 b) $2\times 8$ ladders; an artificial width of 
	 0.1 is included.}
 \label{figone}
\end{figure}

\begin{figure}[h]
 \null
 \vskip0.8in
 \includegraphics{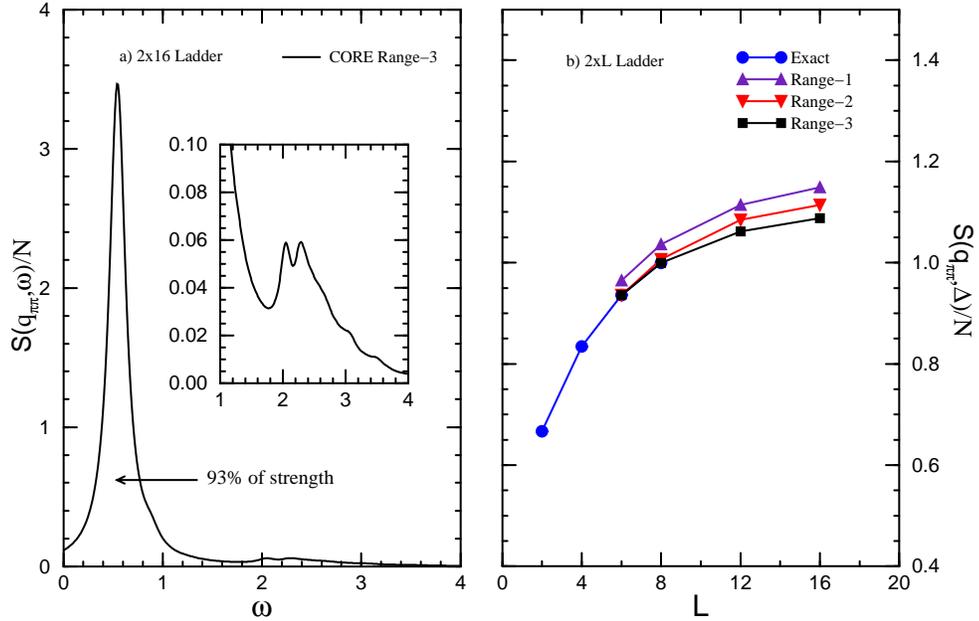}
 \vskip2.6in
\caption{Dynamic spin response a) for a $2\times 16$ ladder; 
	 an artificial width of 0.1 is included and b) to the 
	 lowest triplet state as a function of the length of 
	 the ladder.}         
 \label{figtwo}
\end{figure}

 Because the $\vec S(\vec q_{\pi\pi})$ operator cannot connect states
in the frozen basis to outside states, a high-quality description of 
$S(\vec q_{\pi\pi},\omega)$ utilizing the frozen basis should be 
possible. Moreover, CORE prescribes how any operator---not just the 
Hamiltonian---should be renormalized to account for truncations. We 
here present calculations of $S(\vec q_{\pi\pi},\omega)$ in the frozen 
basis including range-3 renormalization corrections for both the 
Hamiltonian and the transition operator 
for $2 \times L$ systems with $L=6$, $L=8$, and $16$. For $L=6$ a 
range-3 calculation is exact---at least for the low-energy part of 
the response (see Fig.\ref{figone}). For $L=8$ a range-3 CORE calculation 
is no longer exact but we can compare with exact results. We find that 
the range-3 CORE calculation for the integrated response to the 
lowest triplet level, 
$S(\vec q_{\pi\pi},\Delta)\equiv
	|\langle \Psi_{n=1}|\vec{S}(\vec q)|\Psi_{0}\rangle|^2$, 
differs from the exact value by less than 0.014\%. All our calculations 
of $S(\vec q_{\pi\pi},\Delta)$ are summarized in Table~\ref{tableone} 
and in Figure~\ref{figtwo}. There we see that the range-3 renormalization 
corrections to the spin operator [Eq.~(\ref{spipip})]
tend to reduce $S(\vec q_{\pi\pi},\Delta)$; this is true of all induced 
CORE contributions and we therefore assume that the CORE results for the 
$2 \times 16$ ladder are upper limits. Based on the accuracy of CORE for 
$2 \times 8$ ladders, we speculate that the error in our $2 \times 16$ 
result is less than 0.1\%. 

 The following picture of $S(\vec q_{\pi\pi},\omega)$ emerges from our 
calculations. As the length of the ladder grows from $L=2$, the great 
majority of strength remains concentrated in the lowest triplet state.
At the same time, moderately strong coherent many-body effects increase 
the response/site to the first triplet state from 2/3 for the smallest 
system for which this response can be defined, namely, a single 
$2 \times 2$ plaquette, to a value of 1.088 for $L=16$. Extrapolation 
to the thermodynamic limit is somewhat ambiguous but we speculate that, 
in this limit, $S(\vec q_{\pi\pi},\Delta)$/site will lie between 1.10 
and 1.12---which corresponds to an enhancement over the single-plaquette 
value by a factor of around 1.6. It would be very interesting to see 
the extent to which these predictions are consistent with, {\it e.g.}, 
future neutron scattering measurements utilizing single-crystal ladder 
compounds.

 In summary, we have used the recently proposed plaquette 
basis~\cite{piek96} to investigate static and dynamic properties of 
isotropic 2-leg Heisenberg spin ladders. This basis reveals some of 
the important physics of these complicated systems in an especially 
clear fashion. For example, we find that the low-energy states of 
these ladders are dominated by the relatively few configurations
in which the diagonal spins of the plaquettes are frozen in triplets. 
This means that an extreme truncation of the basis which retains only 
the frozen states will nevertheless be reasonably accurate. When 
truncation errors are corrected for using the recently developed 
CORE approach~\cite{morwei94}, we are able to compute properties 
of $2 \times 16$ ladders with great precision. In addition to 
static quantities such as the energy/site and the gap, we are also 
able to calculate the dynamical spin response 
$S(\vec q_{\pi\pi},\omega)$. As the operator which governs this 
response cannot break the frozen spin triplets, accurate calculations 
of $S(\vec q_{\pi\pi},\omega)$ are possible in the truncated basis 
and these, too, can be systematically improved via the CORE technology. 
Moreover, just as a simple picture of non-interacting multi-plaquette 
states gives a surprisingly accurate estimate of the energy/site of 
large systems, the same picture also tells us a great deal about 
$S(\vec q_{\pi\pi},\omega)$. Specifically, we see that the bulk
of the response is concentrated in the lowest triplet state 
and that the response/site is of order 1. 

We emphasize the computational efficiencies achievable when using 
CORE with the truncated plaquette basis. Because the bulk of the 
$S(\vec q_{\pi\pi},\omega)$ response lies in the lowest triplet 
excitation and we can focus on this single transition, still greater 
efficiencies are possible. Specifically we may use a simple 
$2 \times 2$ Lanczos technique~\cite{lancz2} to find only the 
lowest eigenvalue and eigenvector in the singlet and triplet sectors
rather than performing the much more time-consuming full 
diagonalizations. With these efficiencies, calculation of the 
$2 \times 16$ response $S(\vec q_{\pi\pi},\Delta)$ presented here
required only 12 minutes of CPU time on a PC. In addition to 
providing new results for the dynamic spin response---which we hope 
can soon be compared with new data for inelastic neutron scattering 
from single crystals of ladder compounds---we have demonstrated the 
utility of the plaquette basis both for numerical studies and for 
illuminating the important physics of spin ladders.

\acknowledgments
We thank E. Dagotto, M. Laukamp, F.X. Lee, and G.B. Martins 
for many useful and stimulating discussions. This work was 
supported by the DOE under Contracts Nos. DE-FC05-85ER250000, 
DE-FG05-92ER40750 and DE-FG03-93ER40774.

\mediumtext
 \begin{table}
  \caption{Static and dynamical properties of a $2\times 8$ 
	   (first set of numbers) and a $2 \times 16$ (last
	   set of numbers) Heisenberg ladder. Quantities in 
   	   parenthesis represent the discrepancy with the exact 
	   or DMRG values, respectively.} 
   \begin{tabular}{c|cc|cc}
   Observable    & CORE range-3    &  Exact    
                 & CORE range-3    &  DMRG    \\
     \tableline
   $E_{0}$/site  &$-0.55711(0.020\%)$&$-0.55722$   
                 &$-0.56742(0.037\%)$&$-0.56763$   \\
   $E_{1}$/site  &$-0.51784(0.012\%)$&$-0.51778$   
                 &$-0.55041(0.018\%)$&$-0.55031$   \\
   $E_{1}-E_{0}$ &$\quad 0.62835(0.423\%)$&$\quad 0.63101$    
                 &$\quad 0.54441(1.782\%)$&$\quad 0.55411$   \\
   $S(\vec q_{\pi\pi},\Delta)$/site   &$\quad 0.99920(0.014\%)$ 
                                      &$\quad 0.99934$    
  		     &$\quad 1.08785\phantom{(0.014\%)}$ & NA \\ 
   \end{tabular}
  \label{tableone}
 \end{table}

\end{document}